\documentclass[preprintesaps,twocolumn,floatfix,superscriptaddress]{revtex4}
\usepackage{graphicx}
\usepackage{dcolumn}
\usepackage[rgb,dvipsnames]{xcolor}
\usepackage{latexsym}
\usepackage{amsmath, amsthm, amssymb}
\usepackage{epsfig}
\usepackage{latexsym}
\usepackage{bm}

\newcommand{\eq}[1]{Eq.~(\ref{#1})}
\def\mean#1{\left< #1 \right>}

\begin{document}

\title{Synchronization in an evolving network}

\author{R K Singh}
\email[]{rksingh@imsc.res.in}
\affiliation{The Institute of Mathematical Sciences, CIT Campus, Taramani, Chennai 600113, India}

\author{Trilochan Bagarti}
\email[]{bagarti@hri.res.in}
\affiliation{Harish-Chandra Research Institute, Chhatnag Raod, Jhunsi, Allahabad 211019, India}

\begin{abstract}
In this work we study the dynamics of Kuramoto oscillators on a stochastically evolving network whose evolution is 
governed by the phases of the individual oscillators and degree distribution. Synchronization is achieved after a threshold 
connection density is reached. This cumulative effect of topology and dynamics has many real-world implications, where 
synchronization in a system emerges as a collective property of its components in a self-organizing manner. The synchronous state remains 
stable as long as the connection density remains above the threshold value, with additional links providing resilience 
against network fluctuations. 
\end{abstract}

\maketitle

Synchronization is a ubiquitous phenomenon observed across many natural and artificial systems. Flashing of fireflies, chirping of 
crickets, neurons in brain and Josephson junction arrays are well known examples that show synchronization. Kuramoto model provides a 
paradigm of synchronization of globally coupled phase oscillators \cite{kuramoto,strog,acebron}. However, in a system composed of a 
large number of components, no two of which have identical dynamics, synchronization emerges collectively in a self-organizing manner 
 from a complex web of interactions amongst individual components \cite{kurths}. Dynamics on complex networks provides a manifestation 
for such systems \cite{strogatzNat}. In recent years a number of authors have studied the Kuramoto model on complex networks 
\cite{strogatz}. Past studies have focused on static networks \cite{moreno} but a surge of interest in dynamic networks \cite{saramaki} 
has led people to study the emergence of synchronization in dynamic networks \cite{explosivesync}. In these studies, the dynamics of 
the individual oscillators and the network topology has been treated independently. Synchronization emerges autonomously in many real 
systems which are constantly evolving in time. In this paper, we address the basic question of emergence of synchronization by providing 
a mechanism for the stochastic evolution of the network governed by the system dynamics. Starting initially with a collection of 
independent oscillators we numerically show emergence of global synchronization from an interplay of network topology and local dynamics. 
We also find the synchronous state to be stable as long as the number of nearest neighbors remains above a threshold with the additional 
links providing resilience against network fluctuations \cite{attacks}. 

\label{sec.model}
We consider a system of $N$ coupled phase oscillators described by the following equations : 
\begin{align}
\label{deq1}
\frac{d\theta_{i}}{dt} = \omega_{i} + \frac{\epsilon}{\kappa_i}\sum^N_{j=1} g_{ij}\sin(\theta_j - \theta_i),~~i = 1,\dots,N,
\end{align}
where $\theta_i(t)$ and $\omega_i$ denote the phase and frequency of the \(i^{th}\) oscillator respectively, and $\epsilon$ is 
the coupling strength. Coupling between oscillators is described by the adjacency matrix, ${\bf g} = [g_{ij}]$. For the $i^{th}$ 
oscillator, $\kappa_i = \text{max}\{1,k^+_i\}$, where $k^+_i = \sum_j g_{ij}$, is the in-degree of the node $i$ that counts the number of 
oscillators coupled to the $i^{th}$ oscillator. Similarly, out-degree, $k^-_i = \sum_j g_{ji}$ is the number of oscillators to 
which the $i^{th}$ oscillator is connected. The initial conditions are given by $\theta_i(0) = \theta_{i,0}$ for all $i=1,\ldots,N$ 
and ${\bf g}(0)={\bf g}_0$.

Evolution of the adjacency  matrix ${\bf g}$ is governed by the dynamical rule : 
\begin{align}
\label{dynrule1}
g_{ij}(t_{n+1}) &= 1-g_{ij}(t_{n}) {\text{~with probability~}} P, \nonumber \\
t_{n+1} &= t_{n}+ \tau_{n}, ~~n=0,1,2,\ldots,
\end{align}
where $P$ depends on $g_{ij},~k^\pm_i,~k^\pm_j,~\theta_i,~\theta_j$ for $i \ne j$, and $\tau_{n}$ are continuous random variables. 
From \eq{dynrule1} we observe that the event of a single-flip in adjacency matrix is independent of previous flips (where a flip is 
defined as a change in value of $g_{ij}$ from 0 to 1 and vice-versa). We make a further simplification by assuming that the flips 
occur at a constant rate. Hence, we choose $\tau_{n}$ to be exponentially distributed waiting time intervals of parameter $\lambda$. 
Dynamics of the system proceeds in the following two distinct steps: (a) in a time interval  $[t_n, t_{n+1})$, 
oscillators evolve according to \eq{deq1} with adjacency matrix ${\bf g}(t)={\bf g}(t_n)$ and (b) at time 
$t=t_{n+1}$ a pair of oscillators $(i,j)$ chosen randomly are coupled or decoupled instantaneously by the dynamical 
rule in \eq{dynrule1}. Note that due to stochastic evolution of the network by \eq{dynrule1}, $g_{ij}$ need not be symmetric. 
The mean waiting time $1/\lambda$ and the average time period $2\pi/\mean{\omega}$ provide two natural timescales of 
the system, where $\mean{\omega}$ is the average of the frequency distribution. 

Two special cases of \eq{deq1} and (\ref{dynrule1}) are : 
(a) For $g_{ij} = 1 ~\text{forall}~i\neq j, \text{and}~ 0~\text{otherwise}$, and $\lambda = 0$ we obtain the original 
Kuramoto model \cite{kuramoto}. (b) When the mean waiting time, $\lambda$ is finite and the transition probability $P$ is constant, 
the steady state network is the well known Erdos-Renyi model \cite{barabasi}.

For the general problem defined by \eq{deq1} and (\ref{dynrule1}), the transition probability $P$ for a given pair of oscillators 
$i, j$ depends on the phases of $\theta_i, \theta_j$ and the network degrees of freedom, $g_{ij}, k^\pm_i, k^\pm_j$. We assume that 
the network degrees of freedom are separable from the dynamical variables $\theta_i~\text{and}~\theta_j$. The transition probability 
is defined as : 
\begin{align}
P = (1 - g_{ij})\eta \rho + g_{ij}(1 - \eta)(1 - \rho), 
\end{align}
where $\eta = \eta(k^\pm_i,k^\pm_j)$ and $\rho = \rho(\theta_i,\theta_j)$. 
Here, $\rho$ is an even function of the phase difference $\theta_i - \theta_j$ with a minimum at $\pi$. We choose $\rho$ as : 
\begin{align}
\rho(\theta_i,\theta_j) = (2/\pi)\cos^{2}((\theta_i - \theta_j)/2). 
\label{phrule}
\end{align}

We assume two different functions for $\eta$. Let $\eta_{1(2)} = 1$ for $k^+_{i}=k^-_{j} = 0$ and  
\begin{subequations}
\begin{align}
  \label{rule1}
  \eta_1(k^+_i,k^-_j) &= k^-_j/(k^+_i + k^-_j), \\
  \label{rule2}
  \eta_2(k^+_i,k^-_j) &= 1/(k^+_i k^-_j), 
\end{align} 
\end{subequations}
for $k^+_{i},~k^-_{j}$ nonzero. \eq{rule1} is motivated by the observation that hubs in a network are generally more influential 
as compared to less interacting nodes \cite{scalefree}. \eq{rule2} on the other hand captures the form of dynamics where the tendency 
of a node to get connected to more nodes diminishes with its increasing neighbors \cite{redner}. 

\label{sec.results}
Initially, all the oscillators are completely isolated (i.e. ${\bf g}(0) = {\bf 0}$) with the phases and frequencies distributed 
uniformly in the interval $(0, 2\pi)$ and $(\pi - 1/2, \pi + 1/2)$ respectively. Average phase of the oscillators is defined as : 
\begin{align}
\label{phase}
r e^{i \psi} = \frac{1}{N}\sum^N_{j=1} e^{i \theta_j}, 
\end{align}
where $\theta_j$ is the phase associated with oscillator $j$ and $\psi$ the average global phase. The magnitude $r$ is a measure of 
the global synchronization of the system. For a completely phase-locked state, $r = 1$. The connection density of the 
network is defined as : 
\begin{align}
\label{connection}
c = \frac{E}{N(N - 1)},
\end{align}
where $E$ is the number of directed links in the graph at any given instant and $N$ the total number of nodes. $c$ 
is a measure of the denseness of the graph. For a completely connected graph in which every node is connected to 
every other node, $c = 1$. 

The quantities of interest defined above, we proceed to solve the set of equations in \eq{deq1} coupled with \eq{dynrule1} in 
order to study the evolution of the system as a whole. Numerical calculations suggest the existence of a critical value of global 
coupling strength, $\epsilon_0$, below which the system remains in an asynchronous state but above which all the oscillators 
phase-lock with each other. We analyse the properties of the system for a value of coupling strength $\epsilon > \epsilon_0$ and 
then calculate the value of $\epsilon_0$ using finite size scaling \cite{barkema}. 
\begin{figure}
\includegraphics[width=0.5\textwidth]{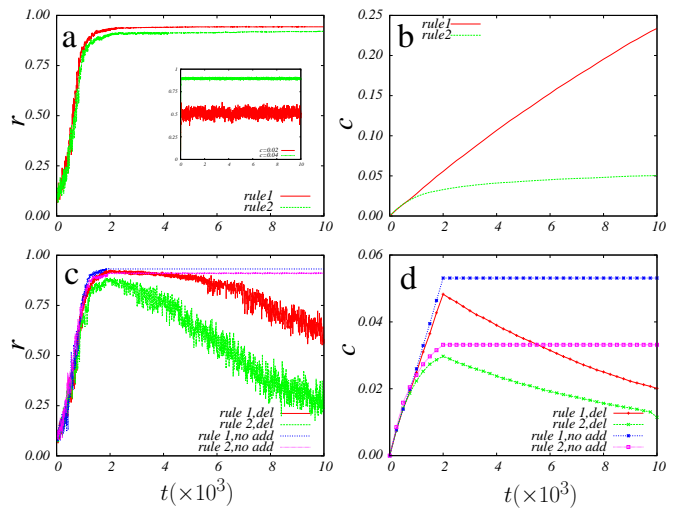}
\caption{(Color online) Variation of $r$ and $c$ with $t$ for the given initial conditions with $\epsilon = 1,~N = 100,~\lambda = 1$ : 
$(a)$ The system gets synchronized in almost the same time for both the transition rules \eq{rule1} and \eq{rule2}. The inset shows 
the rise in order parameter for a static network topology at $c = 0.02,~0.04$. $(b)$ $c$ grows sublinearly with $t$, with the rate 
of increment different for both the rules. $(c)$ Onset of synchronization in the system and its persistence after the network is not 
allowed to change any further once the system achieves a global phase-locking. If the links are removed randomly from the network then 
the synchronization is lost. $(d)$ Connection density for the behavior in $(c)$. The results are obtained after averaging over 10 
ensembles. }
\label{resilience}
\end{figure}

Fig.~\ref{resilience} shows the variation of order-parameter, $r$ and connection, $c$ with time, $t$ for global coupling strength, 
$\epsilon = 1$, for a system of size $N = 100$ with the parameter of the waiting time distribution, $\lambda = 1$. As observed from 
Fig.~\ref{resilience}(a),(inset), it takes a comparatively longer time for the system of phase oscillators to reach a 
phase-locked state in an evolving network as compared to a network of static topology with a given connection density. 
Sublinear growth of $c$ with $t$ is shown in Fig.~\ref{resilience}(b). We have evolved the system upto $t = 50000$ and 
observed that the connection density $c \sim t^a, a < 1$ for both the transition rules. The exact value of exponent $a$ depends on the transition rules 
and the value of coupling strength. The time series for $r$ in the two figures also suggests towards 
existence of a threshold value of connection density in order to achieve global synchronization. 
To confirm the existence of a threshold value of connection density, we stop adding new links to the network once the system 
achieves global synchronization. We observe that the synchronization persists thereafter providing a strong indication towards the 
existence of a minimum number of oscillators which should interact so that the system evolves coherently as a whole. Once the 
network stops evolving, we check the robustness of the synchronized state and find that synchronization is not resilient to 
random removal of links as the network gets fragmented(see Fig.~\ref{resilience}(c),(d)). 

\begin{figure}
\includegraphics[width=0.5\textwidth]{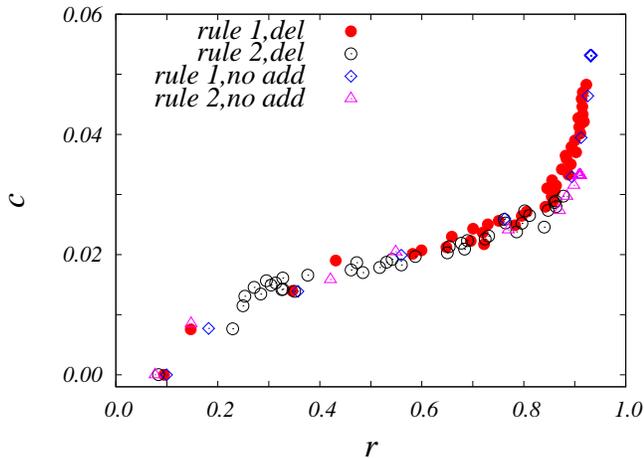}
\caption{(Color online) Phase diagram of connection, $c$ vs order-parameter, $r$. Once the synchronization is achieved, 
$c$ rises abruptly due to rise in network density, but more addition of links has a negligible effect on the value of order parameter. }
\label{phasediag}
\end{figure}

In Fig.~\ref{phasediag} is the phase diagram $r$ vs $c$ show a monotonic increase of $c$ with $r$. We observe that a threshold value 
of $c$ is required in order to achieve synchronization. Once the synchronization is achieved, $r$ becomes 
constant but $c$ rises due to addition of links on an average. Existence of a threshold value of $c$ can be understood from the 
observation that size of the largest connected component of the directed graph quickly becomes $N$, the system size. This is 
achieved well before synchronization sets in. This also explains the desynchronization in the system due to random removal of 
links which makes the network fragmented into many components. This observation was confirmed by walking on the graph using depth 
first search algorithm, which provides the size of the connected components in a graph \cite{dfsref}. A threshold value of $c$ 
required for the system to achieve synchronization also suggests that a system of larger size will take longer time to achieve 
synchronization in comparison to a system of smaller size which is confirmed numerically. Numerically, exact value of threshold 
connection depends on the value of the order-parameter at which the system is considered to be synchronized, e.g., if 
$r \approx 0.9$ is chosen to be the value of order-parameter for global synchronization, the corresponding threshold connection 
density is, $c \approx 0.03$. 

\begin{figure}
\includegraphics[width=0.5\textwidth]{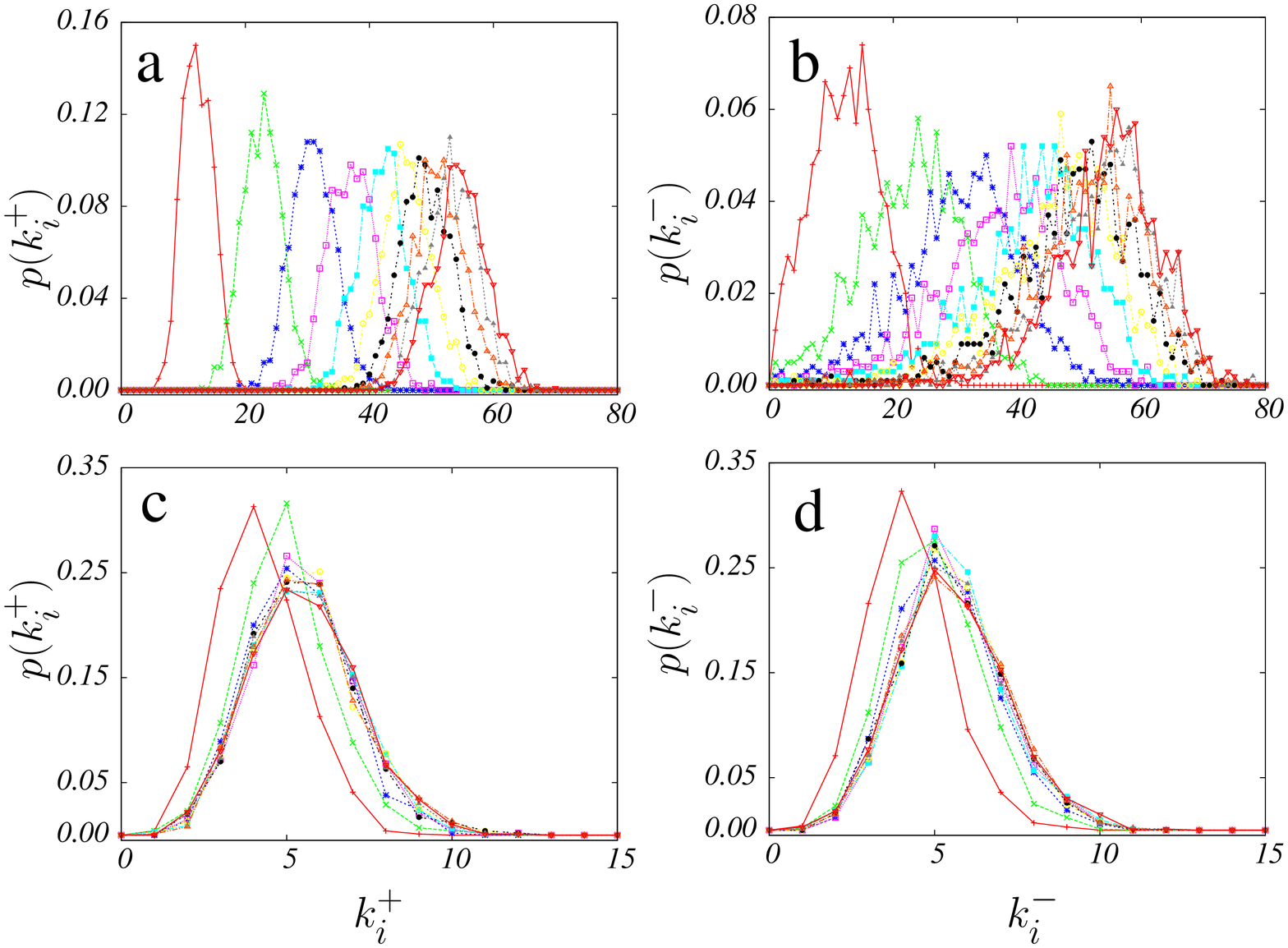}
\caption{(Color online) Degree distribution for the evolving network with $N = 100$, 
$\epsilon = 1$ and $\lambda = 1$. (a) and (b) are the in-degree and out-degree 
distribution for the first transition rule of network evolution. Likewise (c) and (d) are for the second transition rule. 
The system was evolved upto $t = 50000$ and the distribution curves are from $t = 5000$ to $t = 50000$ in steps of 
$\Delta t = 5000$ with the curves shifting towards right in time. The results are obtained after averaging over 10 ensembles. }
\label{degdist}
\end{figure}

Fig.~\ref{degdist} shows the evolution of degree distribution with time for the network. The mean of the degree distribution 
$\mean{k^\pm}$ grows sublinearly in time with $\mean{k^+} = \mean{k^-}$ for both the transition rules. The equality of mean 
degree is evident from the fact that total in-degree equals total out-degree for a directed network and hence, every node is 
affected almost equally by its neighbors. Variances on the other hand are not identical for both the in- and out-degree distributions 
for rule 1, with the in-degree distribution showing less width in comparison to the out-degree distribution(Fig.~\ref{degdist}(a),(b)). 
Note that for rule 1, $\eta_1(k^+_i,k^-_j) \neq \eta_1(k^-_j,k^+_i)$, and since $\eta_1$ depends on the ratio of 
the out-degree of node $j$ to the sum of in-degree of node $i$ and out-degree of node $j$. As a result different oscillators contribute 
differently towards affecting the dynamics of a given oscillator. However, for the second transition rule, we see that 
$\eta_2(k^+_i,k^-_j) = \eta_2(k^-_j,k^+_i)$, therefore the width is identical for both the in- and out-degree 
distributions for obvious reasons. 

\begin{figure}
\includegraphics[width=0.5\textwidth]{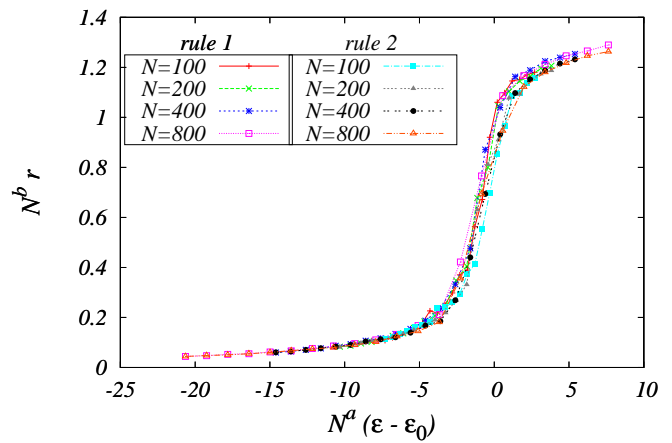}
\caption{(Color online) Finite size scaling at mean-waiting time, $\lambda = 1$ of $r$ vs $\epsilon$ curve for different system 
sizes, $N$ for the two transition rules. The values of the critical exponents $a$ and $b$ are : $a \approx 0.5,\; b \approx 0.05$. 
The critical value global coupling strength required to set the system of phase oscillators in a synchronous state is $\epsilon_0 
\approx 0.73$. The results are obtained once the system reaches steady state. The calculations are done in $t \rightarrow \infty$ 
limit and averaged over 10 ensembles. }
\label{fss}
\end{figure}

Finite size scaling of the transition curve(Fig.~\ref{fss}) shows that the properties observed from numerical solution of \eq{deq1} 
do not depend on system size. Finite size scaling analysis also provides an approximate value of the critical coupling strength, 
$\epsilon_0$ below which the system will never reach a phase locked state independent of its evolution time. The value of 
critical coupling strength is the same for both the transition rules and this can be understood in terms of the connection threshold. 
This is because once the system reaches synchronization, increasing connection density further does not have any significant effect on 
the order parameter. Furthermore, a finite value of critical coupling strength is also evident from the observation 
that the second moment of the degree distribution of the steady-state network is finite \cite{variance}. 

We benchmark our results against existing literature by solving the dynamical equations with oscillator frequencies 
distributed according to Cauchy distribution peaked at the origin and having scale parameter, $\gamma = 0.25$. In the mean field 
limit for a completely connected graph, the critical value of coupling strength is, $\epsilon_0 = 2 \gamma$ \cite{acebron}. We 
evolve the network by adding links with probability one to randomly chosen pair of nodes so that the steady state topology of the 
network is that of a completely connected graph. We find $\epsilon_0 \approx 0.4$, which provides a benchmark to our results. 

In order to investigate the dependence of $\epsilon_0$ on the mean of the waiting time distribution $\lambda$, we looked at the 
variation of order-parameter with coupling for different values of $\lambda$ for a fixed system size. We find 
that varying $\lambda$ does not have any appreciable effect on the value of critical coupling strength for either of 
the rules. This observation suggests that $\epsilon_0$ is independent of $\lambda$. 

To conclude, we have shown that the interdependence of topology and dynamics is necessary in order to understand how synchronization 
emerges collectively from local interactions in a self-organizing manner. This has real world implications for systems in which 
interactions keep changing with time but an overall rythm persists on an average. It is also observed in natural systems that the 
synchronous state does not persist forever. In this model we have provided a mechanism for collective emergence of synchronization 
from local dynamics. The synchronization achieved is characteristic of the network evolved by the transition rules. However, it is 
independent of the connection density of the network provided it is above a threshold. Existence of a connection density threshold 
in order to achieve synchronization has practical applications, e.g., from the context of Josephson junction arrays in reducing the 
number of redundant connection links in order to achieve desired output thereby minimizing the cost of wirings. We hope that this 
work provides an impetus for future work on synchronization from a dynamic perspective.

\end{document}